\documentclass[reprint,superscriptaddress,amsmath,amssymb,aps,floatfix]{revtex4-2}

\usepackage{graphicx}
\graphicspath{{images/}}
\usepackage{dcolumn}
\usepackage{bm}
\usepackage{xr-hyper}
\usepackage{hyperref}
\usepackage{xcolor}

\begin{document}

\preprint{APS/123-QED}

\title{Temporal shaping of narrow-band picosecond pulses via non-colinear sum-frequency mixing of dispersion-controlled pulses}

\author{Randy Lemons}
 \email{rlemons@slac.stanford.edu}
\affiliation{%
 ~SLAC National Accelerator Laboratory and Stanford University, 2575 Sand Hill Rd, Menlo Park, CA 94025
}%
\affiliation{%
 ~Colorado School of Mines, 1500 Illinois St, Golden, CO 80401
}%
\author{Nicole Neveu}
 \email{nneveu@slac.stanford.edu}
\author{Joseph Duris}
\author{Agostino Marinelli}\affiliation{%
 ~SLAC National Accelerator Laboratory and Stanford University, 2575 Sand Hill Rd, Menlo Park, CA 94025
}%
\author{Charles Durfee}
\affiliation{%
 ~Colorado School of Mines, 1500 Illinois St, Golden, CO 80401
}%
\author{Sergio Carbajo}
 \email{scarbajo@slac.stanford.edu}
\affiliation{%
 ~SLAC National Accelerator Laboratory and Stanford University, 2575 Sand Hill Rd, Menlo Park, CA 94025
}%

\date{\today}

\begin{abstract}
A long sought-after goal for photocathode electron sources has been to improve performance by temporally shaping the incident exitation laser pulse. The narrow bandwidth, short wavelength, and picosecond pulse duration make it challenging to employ conventional spectral pulse shaping techniques. We present a novel and efficient intensity-envelope shaping technique achieved during nonlinear upconversion through opposite-chirp sum-frequency mixing. We also present a numerical case-study of the LCLS-II photoinjector where transverse electron emittance is improved by at least 25\%.

\end{abstract}

\maketitle


\section{Introduction}

Generation of picosecond optical laser pulses with tailored temporal intensity envelopes is a technically challenging task. Their narrow spectral bandwidth restricts methods that attempt to modify temporal intensity through spectral phase and their temporal duration is too short to exploit direct electronic modulation methods. However, picosecond duration pulses and shaping their temporal intensity profile is critically important for photoinjector-based electron instruments. Photoinjectors lie at the heart of a myriad of forefront scientific research areas and technologies in quantum electrodynamics~\cite{yakimenko2019prospect,wang2020coherent}, high-energy physics~\cite{ellis2001new,leemans2006gev}, ultrafast science~\cite{emma2010first,kern2018structures,stankus2019ultrafast,ibrahim2020untangling}, and medical technologies~\cite{neutze2000potential,coquelle2018chromophore,wolf2019photochemical} among many others. Even though the operational requirements for applications dependent on photoinjectors can be vastly different, increased electron control and performance remains a central challenge as these various technologies are brought to new scientific and engineering frontiers. From large scale facilities such as storage rings~\cite{bilderback2005review}, electron-electron colliders~\cite{ellis2001new}, and x-ray free electron lasers (XFELs)~\cite{mcneil2010x}, to laboratory-scale instrumentation such as ultrafast electron diffraction~\cite{hastings2006ultrafast} or transmission electron microscopy~\cite{li2014single}, electron generation in a photoinjector exists as the crucial first step and a significant predictor of instrument performance. One foundational strategy to tailor the electron bunch phase-space for greater uniformity and higher brightness is to control the shape of the optical pulses used for photoemission.


In photoinjectors, electrons are generated via the photoelectric effect with laser pulses comprised of light, typically in the deep ultraviolet (UV), whose photon energy lies above the work function of the cathode material. After photoemission, electrons' phase-space distribution evolves rapidly from the interaction with the external accelerating fields and their own internal coulombic forces~\cite{palmer1998next}. As part of this process, the spatio-temporal distribution of the laser pulses effect the electron distribution, and thereby can be used to influence how the bunches will evolve as they are accelerated in vacuum. A key measure of electron bunch quality is its transverse emittance, $\epsilon_x$, defined as~\cite{Wiedemann2015}
\begin{equation}
    \epsilon_x = \sqrt{\langle x_{i}^{2} \rangle \langle x_{i}^{\prime 2} \rangle - \langle x_{i}^{2} x_{i}^{\prime 2} \rangle \label{EQN:Emittance}}
\end{equation}
where $x$ is transverse position and $x^{\prime}$ is the corresponding angle with respect to the ideal trajectory. Generally, emittance reduction is beneficial to the electron beam brightness and any associated secondary emission processes, and can be therefore be used as a metric to optimize in determining spatio-temporal laser distributions~\cite{stephan2020high}.

In XFELs, in addition to low emittance ($<1.5$ $\mu$m), it is particularly crucial to have electron bunches with narrow energy spread ($\Delta \mathrm{E}/\mathrm{E} < 10^{-3}$) and good spatial uniformity, as growth in these parameters can significantly decrease x-ray production in the undulators~\cite{PhysRevSTAB.10.034801}. Beyond conventional Gaussian temporal and spatial distribution of the photoinjector laser pulses, other commonly sought-after laser distributions shown to reduce normalized transverse emittance include flat-top spatio-temporal profiles resembling cylinders~\cite{krasilnikov2012experimentally} or 3D ellipsoids such that the beam size and intensity vary as a function of time~\cite{luiten2004realize}.

Existing shaping techniques rely on modifying or combining femtosecond pulses and can be split into either spectral techniques such as spatial-light modulators~\cite{mironov2016shaping,penco2013optimization} or acousto-optic modulators~\cite{li2009laser,petrarca2007production}, or temporal techniques such as pulse stackers~\cite{krasilnikov2012experimentally,krasilnikov2019studies}. Spectral methods are hindered by tailored phase structure being distorted during pulse amplification and upconversion if shaping in the IR, limited spectral bandwidth if shaping in the UV, and by material damage threshold limitations in the mW-level~\cite{carbajo2018power}. On the other hand, intensity fluctuations inherent to temporal techniques have been shown to induce unwanted microbunching~\cite{PhysRevAccelBeams.23.024401,mitchell2016sensitivity} on the electron bunch, resulting in increased emittance relative to Gaussian distributions. In order to operate at the correct wavelength, these lasers also typically employ a series of nonlinear conversion stages to upconvert infrared (IR) light to UV light below 270 nm~\cite{will2011photoinjector,gilevich2020lcls,alley1999design}. Efficient nonlinear conversion is detrimentally affected by non-flat phase structure and can distort tailored temporal profiles, complicating shaping efforts. As such, these methods are limited in their applicability to high average power, 24/7 facilities, such as LCLS where reliability is paramount.

\begin{figure}[htp!]
	\centering
	\includegraphics[width=1\linewidth]{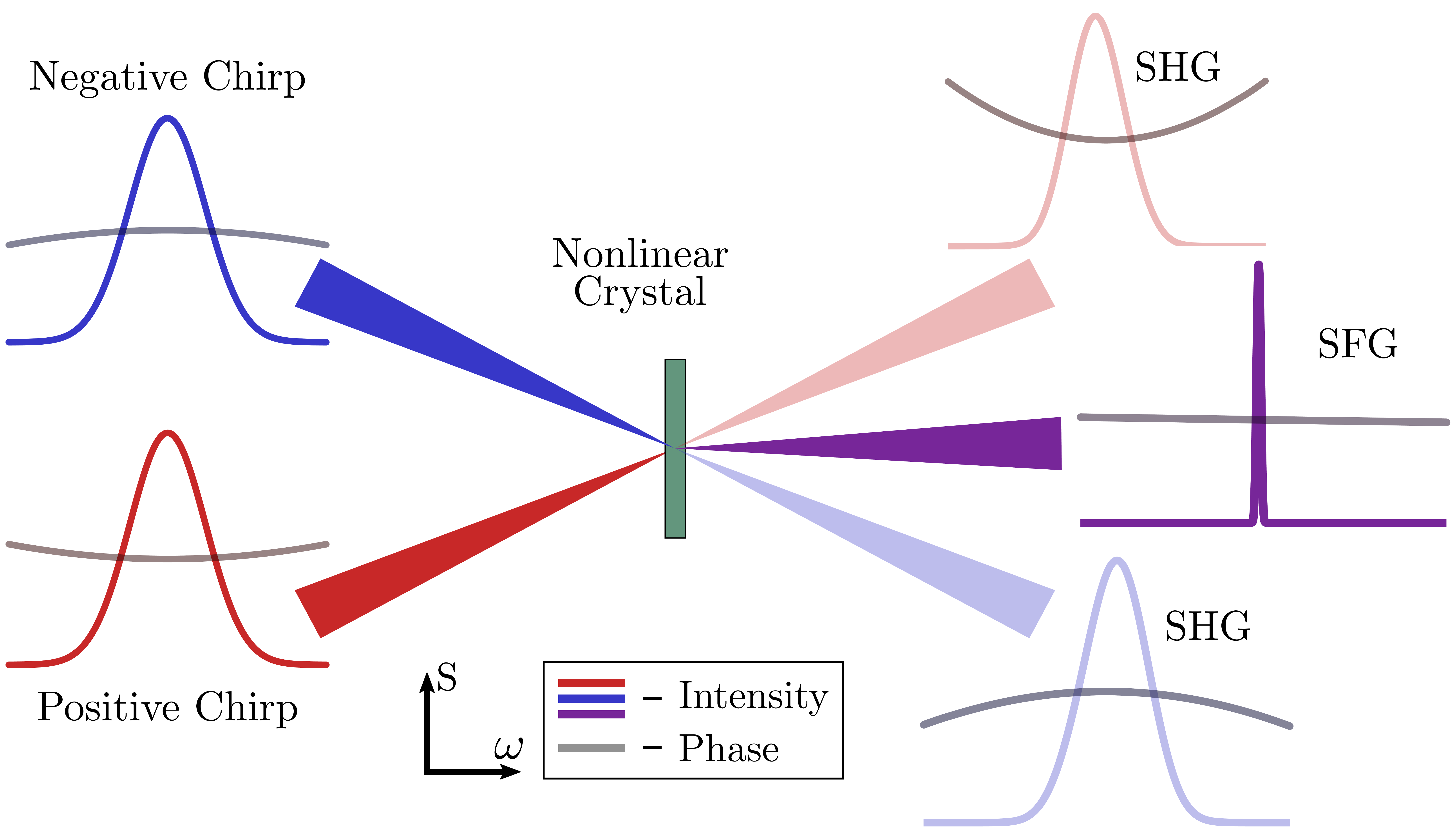}
	\caption{DCNS method in the spectral domain demonstrating two pulses of equal on opposite chirp mixing in a nonlinear medium to generate chirped second harmonic (SHG) copies of each pulse along with the narrowband, flat phase, sum frequency (SFG) pulse.}
  \label{FIG:OpChirp_Schematic}
\end{figure}

We present an upconversion method that incorporates temporal shaping into a nonlinear conversion stage exploiting post-amplification phase manipulation, thereby circumventing the pitfalls of existing upconversion and shaping techniques. We utilize non-colinear sum frequency generation (Fig. \ref{FIG:OpChirp_Schematic}) combining two highly dispersed pulses that result in a pulse with tunable temporal profile in duration and shape~\cite{vicario2012deep,kuzmin2021highly}. Here, we build on the elegant approach by Raoult et al~\cite{raoult1998efficient} of efficient narrowband second harmonic generation in thick crystals by adding higher-order dispersion to simultaneously shape the output pulse duration. The output pulse is nearly transform limited and can be directly utilized for application or passed through further nonlinear conversions processes without distortion. As such, this method, which we call dispersion controlled nonlinear shaping (DCNS), can be broadly used to tailor pulses for the reduction of normalized transverse emittance in photoinjector-based instrumentation.

\section{Dispersion-controlled Pulses}\label{SEC:optics}
To describe DCNS, we start off by expressing the electric field of a laser in frequency space as $E(\omega)=A(\omega)e^{i\varphi(\omega)}$, where $A(\omega)$ is the spectral amplitude and $\varphi(\omega)$ is the spectral phase as a function of angular frequency $\omega$. The amplitude function is typically Gaussian in shape and the spectral phase is customarily described by a Taylor expansion around the central frequency of the field, $\omega_0$.
\begin{multline}
    \varphi(\omega) = \varphi_0 + \varphi_1 (\omega-\omega_0) + \frac{\varphi_2}{2!} (\omega-\omega_0)^2 \\
    + \frac{\varphi_3}{3!} (\omega-\omega_0)^3 + \frac{\varphi_4}{4!} (\omega-\omega_0)^4 + \dots \label{EQN:SpecPhase}
\end{multline}
Here $\varphi_j$ is the $j^{th}$ derivative of $\varphi(\omega)$, evaluated at $\omega_0$.

The first two terms of \eqref{EQN:SpecPhase} correspond to the absolute phase at the central frequency and the group delay, respectively. Since these two terms do not affect the shape of the pulses, our investigation will focus on the next terms, higher order in $\omega$, to control the shape of the pulse envelope. The second order dispersion (SOD) term, $\varphi_2$, is a quadratic phase that gives a linear ordering to the arrival time of the frequency components, stretching the duration of the pulse. Third order dispersion (TOD) results from cubic phase, $\varphi_3$. Here the opposite ends of the spectrum will overlap in time either before or after the main peak, leading to characteristic beating or temporal fringes. Higher orders effect the pulse similarly, but to a lesser degree, with even orders primarily effecting pulse duration and odd orders leading to modulation of the pulse in the time domain. In our description, we will focus on using a combination of SOD and TOD as the two most influential parameters to control the shape of the output pulse in the nonlinear sum frequency mixing process.

\section{Methods \& Results}
We begin by modeling these pulses as a combination of two equal-energy transform-limited Gaussian pulses overlapped in time. From here, the phase of each pulse can be adjusted separately by multiplying with Eq.\ref{EQN:SpecPhase} in frequency space, where $\varphi_2$ and $\varphi_3$ are free parameters and higher-order terms are ignored. These tailored pulses then become our two initial fields, $A_1$ and $A_2$, used in solving the coupled equations (Eq. \eqref{EQN:SFGEqns}) for sum frequency generation~\cite{boyd2019nonlinear}.

\begin{subequations}
    \begin{align}
        \frac{d A_1}{d z} &= \frac{2 i d_{eff} \omega_1^2}{k_1 c^2}A_2^* A_3 e^{-i \Delta k z} \\
        \frac{d A_2}{d z} &= \frac{2 i d_{eff} \omega_2^2}{k_2 c^2}A_1^* A_3 e^{-i \Delta k z} \\
        \frac{d A_3}{d z} &= \frac{2 i d_{eff} \omega_3^2}{k_3 c^2}A_1 A_2 e^{i \Delta k z}
    \end{align}
    \label{EQN:SFGEqns}
\end{subequations}

Propagation and frequency mixing is handled using a symmetrized split-step Fourier method along with a fourth order Runga-Kutta algorithm to solve the coupled nonlinear equations. Nonlinear conversion and nonlinear index effects are handled in the time and position domains while propagation and dispersion through the crystal are handled in the temporal and spatial frequency domain. To ensure accurate results, the resolution of the time, frequency, and spatial grids was increased until further refinement resulted in negligible change to the results.

\begin{figure}[htp!]
	\centering
	\includegraphics[width=1\linewidth]{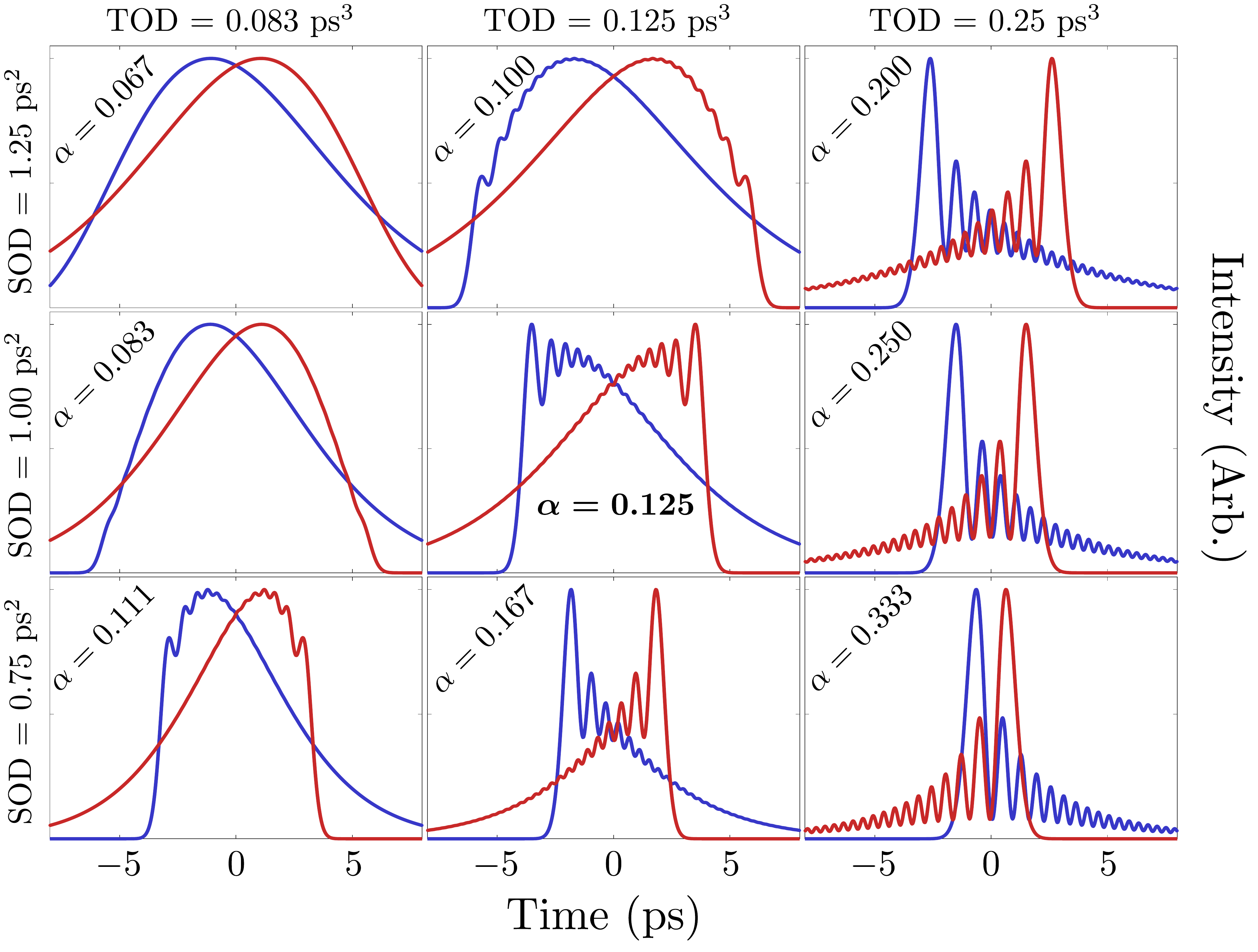}
	\caption{Pulse envelopes resulting from different combinations of SOD and TOD. Each column has a the same magnitude of the TOD value at the top and each row has the same magnitude of SOD. The blue plot results from combinations with negative SOD and positive TOD while the red plot is positive SOD and negative TOD.}
  \label{FIG:3x3_Array}
\end{figure}

As stated above, SOD primarily controls duration and TOD controls the sharpness of the leading or trailing edge and ringing in the field on the opposing edge. It is the interplay between SOD and TOD (Fig. \ref{FIG:3x3_Array}) that then determines the pulse duration and the shape.

By defining the ratio between TOD and SOD, 
\begin{equation*}
    \alpha = \frac{\varphi_3 / ps^3}{\varphi_2/ps^2},
\end{equation*}
we gain a single parameter to describe the general shape of a shaped pulse that is approximately invariant to pulse duration~\footnote{See Supplemental Material at [URL will be inserted by publisher] for reference figure.}. 

Applying these concepts in combination with narrow bandwidth generation suggested by Raoult et al.~\cite{raoult1998efficient} we can search for pulses that fulfill our need for sharp rise time, long pulse duration, and narrow bandwidth.

To best illustrate our conversion scheme and 
to explore how it might be implemented in a real-world case, we use the photoinjector system at LCLS-II. This system consists of an Ytterbium-based commercial laser that outputs 50 $\mu$J, 330~fs pulses with an approximately 4~nm full width at half maximum (FWHM) spectral bandwidth. We propose using type I non-colinear SFG in the conversion from 1030~nm to 515~nm. Using Fig.\ref{FIG:3x3_Array} as a reference to which portion of the pulses will mix, and considering that the LCLS-II design assumed a few 10s of picosecond flat-top photoinjector laser pulse~\cite{osti_1029479}, we focus on situations where the magnitude of $\alpha$ is close to $0.125$~ps (center column). This results in the signal and idler pulses having equal and opposite amounts of SOD and $\alpha=-0.128$~ps, for temporal symmetry. The initial value of SOD ($\approx$3.5 ps$^2$) was chosen so that the FWHM of the 515~nm pulse would be 25~ps in time.

This large amount of SOD necessitates a large amount of TOD to maintain $\alpha$. This can be achieved by passing the transform-limited pulses through a matched compressor-stretcher set with gratings detuned from Littrow angle. To maintain grating efficiency one can generate a larger amount of SOD and an appropriate amount of TOD in the set and reduce SOD to desired levels afterwards with highly dispersive optics, such as chirped volume Bragg gratings~\cite{galvanauskas1998use}.

Once the half-pulse fields are constructed from the given phase parameters, they are propagated with the split-step method through BBO with the crystal angle tuned for type I sum mixing and a 2~mm crystal length. The crossing angle was set to 1.5~deg. This angle must be sufficiently large to allow the sum-frequency signal to be separated from the two input beams and suppress intra-beam second harmonic generation, but not so large as to reduce the spatial overlap of the beams in the crystal~\cite{raoult1998efficient}. Additionally, the simulated focusing through the crystal was adjusted to eliminate back-conversion from 515 nm to 1030 nm as this process can distort the temporal and spectral profiles.

\begin{figure}[htp!]
	\centering
	\includegraphics[width=1\linewidth]{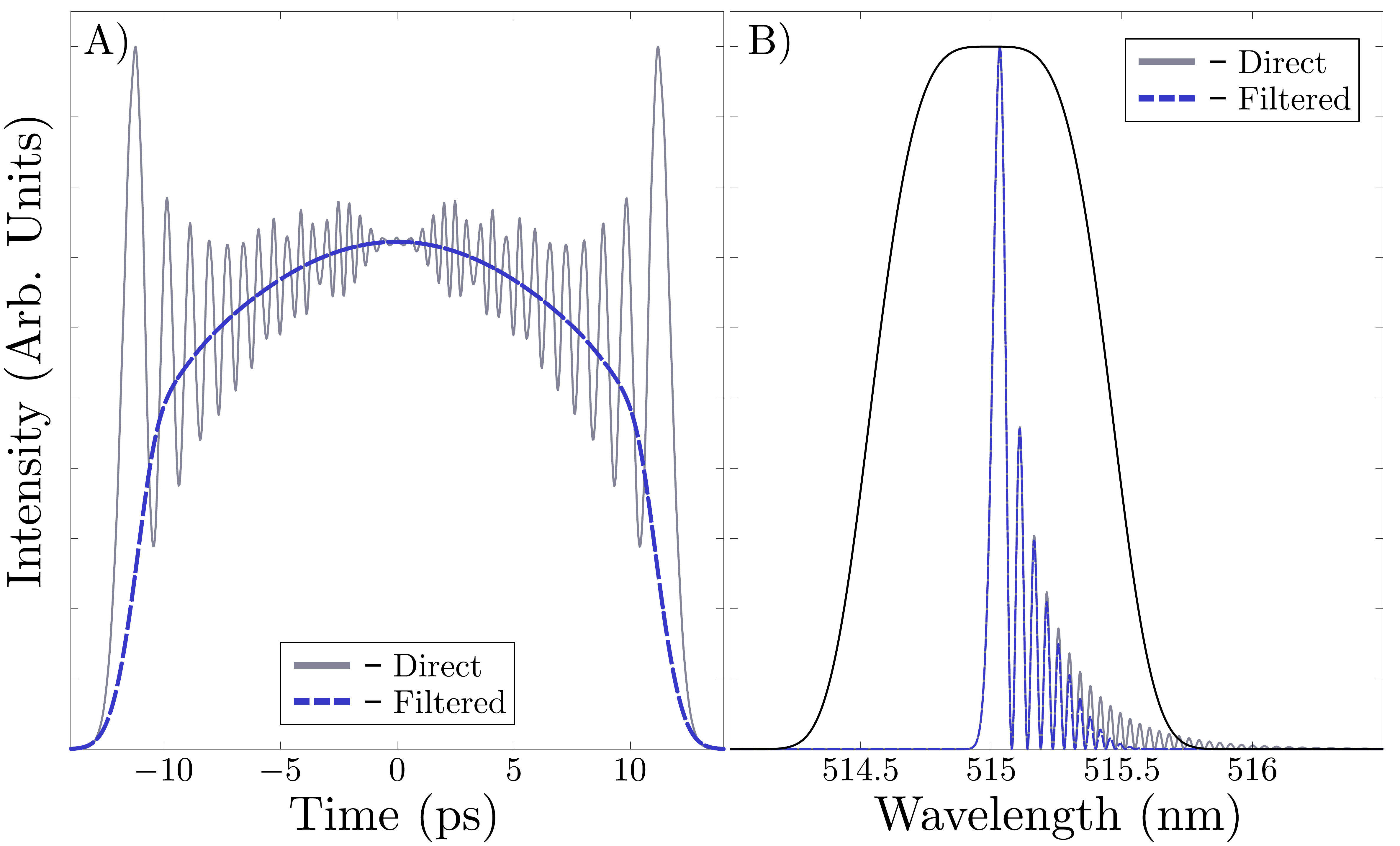}
	\caption{Genetic algorithm optimized 515 nm laser profile for a bunch length of 1.22 mm resulting in an emittance value of 0.30 $\mu$m. Notably, the genetic algorithm settled on a profile which has slower rise and fall times than an ideal flat top as emittance is optimized rather than profile shape. The two plots shown are, a) the laser profile in time before (grey) and after (blue, dashed) a 0.5 nm spectral filter, and b) the spectrum of the pulse before (grey) and after filtering (blue, dashed) with the super-gaussian spectral filter in black.}
  \label{FIG:Fin_Field_Reg_Filt}
\end{figure}


The resultant pulse (Fig. \ref{FIG:Fin_Field_Reg_Filt}a) displays the attractive qualities of a sharp rise time and a flatter profile than the traditionally used Gaussian pulses with upwards of 65\% conversion efficiency in simulation. Experimental efficiency is likely to be less than the 40\% stated by Raoult et al.~\cite{raoult1998efficient} due to added TOD. It is also characterized by large and rapid amplitude fluctuations on the picosecond scale that can be detrimental to e-beam emittance. These fluctuations—the larger oscillations at the edges-—are the result of coherent interaction between the main spectral peak and interference from the higher wavelengths in the asymmetric tail seen in Fig. \ref{FIG:Fin_Field_Reg_Filt}b. However, by applying a spectral amplitude filter after conversion that is significantly wider than the FWHM of the pulse's spectral bandwidth (Fig. \ref{FIG:Fin_Field_Reg_Filt}b), the high frequency components can be attenuated without a major efficiency penalty. In this case, employing a second order super-Gaussian spectral filter with $1/e$ width of 0.5 nm, such as a Bragg grating interference filter, the total power in the field is reduced by less than 10\% and results in a smoother temporal profile (Fig. \ref{FIG:Fin_Field_Reg_Filt}a). As with the mixing, the numerical grids were chosen to reduce possible numerical errors from these filters. Alternatively, the thickness of the nonlinear crystal for the doubling step from 515 nm to 257 nm can be chosen to filter the bandwidth through phase mismatch~\cite{radzewicz2000poor}.

To compare the performance of the DCNS pulses vs. the baseline Gaussian temporal distribution, we simulated the LCLS-II photoinjector performance. To reduce simulation time, the DCNS temporal intensity profiles at 515 nm are squared to directly generate the UV profiles with a root mean squared error of less than 1\%~\footnote{See Supplemental Material at [URL will be inserted by publisher] for quantitative analysis.}. These pulses are supplied as the initial condition to the simulation and used to model the initial electron beam parameters. Electron bunch generation includes variations derived from oscillations in the laser pulse profile. The simulation code used for e-beam dynamics is OPAL~\cite{adelmann2019opal}, and for particle distribution generation, distgen~\cite{distgen}. The FWHM in time and spot size of the laser pulse directly impact the resulting 3D shape of the emitted electron beam. As the beam is emitted, the forces due to external magnetic and electric fields along with internal space charge fields are computed at each time step. In this case, the FWHM is controlled by the magnitude of SOD and shape by $\alpha$. While supplying the DCNS pulses is straightforward, determining the optimal FWHM and spot size of the pulse is not. The strength of the space charge forces are directly impacted by both the FWHM and spot size, which then impacts how strong the external forces need to be to limit emittance growth. To determine optimal laser and machine settings, the parallel simulation code is run in combination with an optimization algorithm (NSGA-II~\cite{dpam:02}) using libEnsemble~\cite{libEnsemble}, a Python library that coordinates ensemble calculations. Standard optimization of the LCLS-II photoinjector includes variables such as laser spot size, laser FWHM, RF cavity phases, RF cavity gradients, and magnet strengths.
To maintain broad applicability of these results, we limit the simulation to only the photoinjector and the first 15 meters of acceleration ($\approx$100 MeV), after which are LCLS-II specific configurations. In this region the laser parameters have the greatest impact on electron bunch evolution as internal space-charge forces are not yet damped by highly relativistic speeds.

\begin{figure}[htp!]
    \centering
    \includegraphics[width=0.45\textwidth]{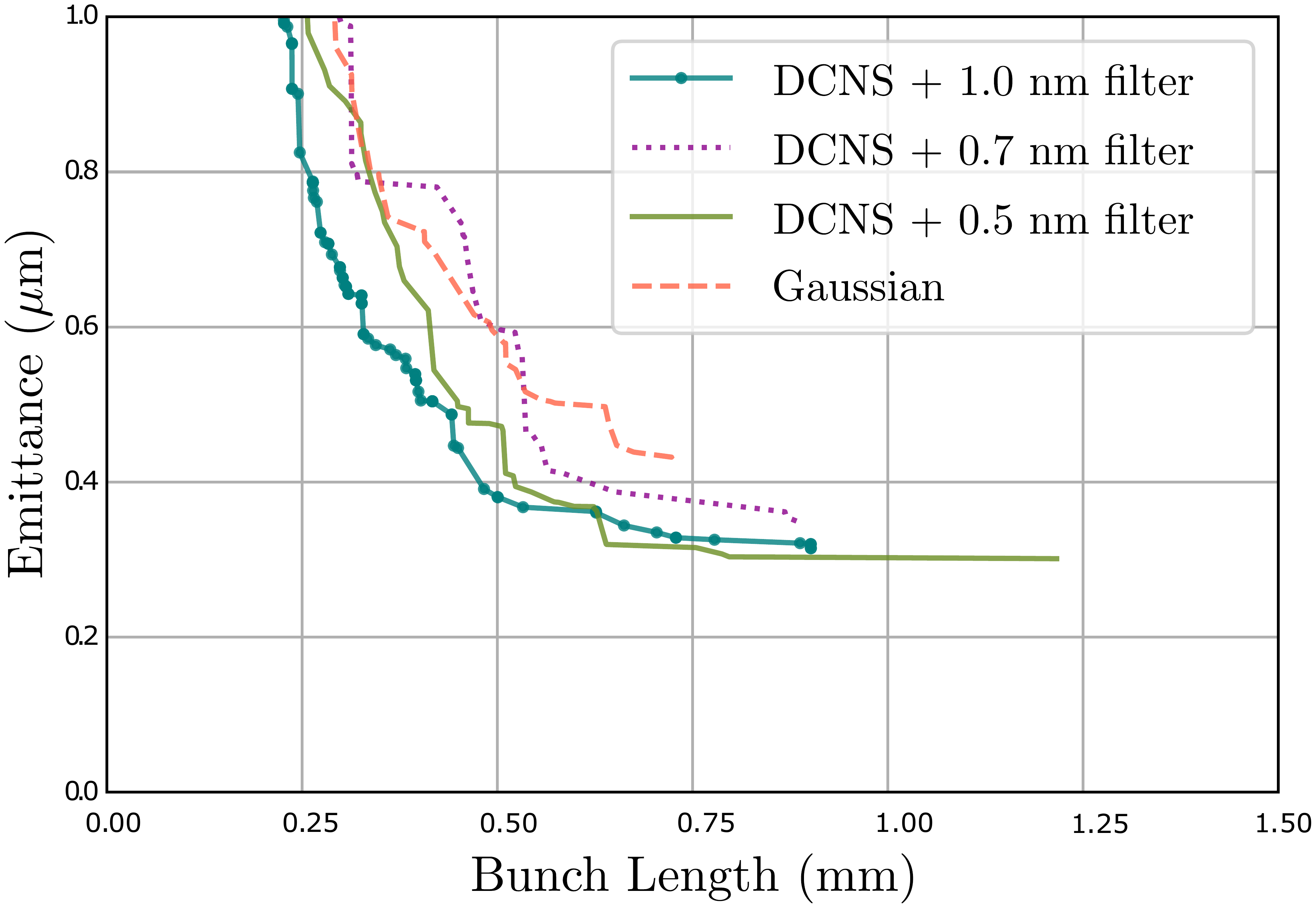}
    \caption{Pareto front comparison of DCNS and Gaussian performance for the LCLS-II injector. DCNS pulses in combination with a 1.0 nm spectral filter, achieves the lowest emittance values at most bunch lengths. The lowest achieved emittance value is 0.30~$\mu$m at a bunch length of 1.22~mm, using a 0.5~nm filter.}
    \label{FIG:optresults}
\end{figure}

The metrics commonly used for determining beam quality for XFELs are emittance \eqref{EQN:Emittance} and bunch length ($\sigma_z$). 
If the emittance is extremely small, the beam size, momentum, or both must be small as well. 
Note, we do not optimize the orthogonal transverse dimension, $y$, because the simulation is transversely symmetric.
As stated earlier, smaller emittances lead to brighter bunches, and therefore better x-ray production.
The same logic applies to second metric of interest, the bunch length, 
which is the rms size of the bunch in the longitudinal dimension. 
 Shorter bunches require less compression after the injector, which mitigates non-linearities in the bunch compression process and results in better FEL performance.

\begin{figure}[htp!]
    \centering
    \includegraphics[width=\linewidth]{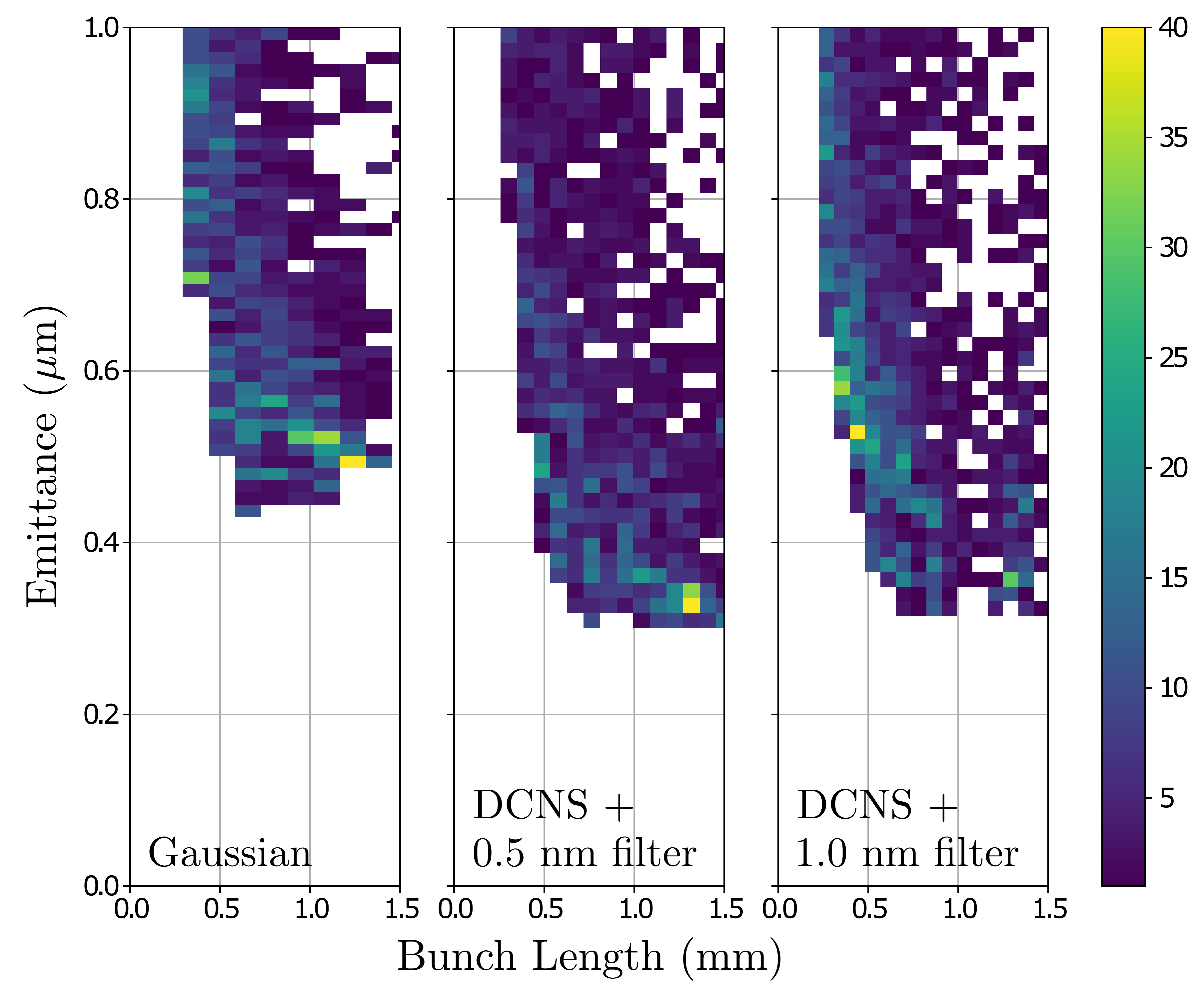}
    \caption{Histogram showing simulation density of all GA solutions, not only Pareto optimal, for Gaussian and DCNS cases. Lighter colored areas indicate a higher number of simulations with valid solutions and thus regions where different methods might be more effective.}
    \label{FIG:hist}
\end{figure}

With these variables and metrics defined, several optimization rounds were performed to compare the performance of DCNS and Gaussian laser pulses in the LCLS-II photoinjector. Final results are shown in Fig.~\ref{FIG:optresults} and Fig.~\ref{FIG:hist}, with the later showing density of simulation points near the Pareto fronts. 
Fig.~\ref{FIG:optresults}, demonstrates that lower emittance values are achieved using DCNS. Further, Fig.~\ref{FIG:hist} indicates the density of low emittance values are shifted for different filters. With a spectral filter of bandwidth 0.5 nm, longer bunch lengths are performing better, whereas with the 1.0 nm filter, there is a higher density of points at shorter bunch lengths. This suggests DNCS could benefit a variety of experimental XFEL configurations with the filter bandwidth and $\alpha$  used as an adjustment knob. 

As shown in Fig.~\ref{FIG:optresults}, the best value found is $\epsilon_x=0.30$~$\mu$m with the the DCNS pulse shown in Fig.~\ref{FIG:Fin_Field_Reg_Filt}.
The bunch length in this case, 1.22 mm, 
is slightly longer than the typical operating length of 1 mm at LCLS.
For a practical comparison, we choose the minimum emittance values at $\sigma_z=1$ mm.
A 25\% improvement in the emittance value is
obtained from DCNS ($\epsilon_x=0.37$~$\mu$m) vs. Gaussian ($\epsilon_x=0.50$~$\mu$m) pulses.
Note, this Gaussian point is not visible in Fig.~\ref{FIG:optresults}, 
because it is not Pareto optimal. 
For shorter bunch lengths, i.e. 0.5 mm, the difference is slightly larger
reaching about 30\% ($\epsilon_x=0.4$~$\mu$m vs 0.58~$\mu$m). Applying this reduction to both $x$ and $y$ planes, since the simulation is symmetric, the total transverse brightness can be more than doubled. In the case of XFELs, this emittance improvement translates to a twofold increase in undulator peak brightness, 25\% shorter x-ray wavelengths, and an upper bound reduction in undulator lengths by 25\% for similar peak currents, which can substantially reduce cost, complexity, and size.

\section{Conclusion}
The regime of picosecond intensity envelope shaping is a challenging task that traditionally lies outside the capabilities of many methods commonly implemented for optical pulses with more spectral bandwidth or longer duration. Nonetheless, photoinjector-based electron sources, which would benefit from shaping in this regime, driving understanding of sub-atomic interactions, molecular structure, biological processes, and fundamental particle science utilize known sub-optimal techniques. By tailoring the photoinjector drive lasers, electron emittance, and by extension the electron beam brightness, can be improved enabling further exploration of these research areas and the possibility for broad impact across multiple disciplines.
Existing shaping techniques for the up-converted lasers used in the photoinjectors suffer from challenges in maintaining favorable pulse shapes, providing enough photon throughput for excitation, or even increasing electron emittance. Our proposed optical shaping and conversion technique, DCNS, circumvents these issues by directly upconverting optical pulses where the production of favorable temporal distributions is embedded in sum frequency conversion using highly dispersed pulses. In the case of linear accelerators and XFELs such the LCLS-II, this simple solution is expected to improve electron emittance across all investigated bunch lengths over conventional Gaussian pulses with an upwards of 30\% emittance reduction at short bunch lengths (0.25 mm) and 25\% at bunch lengths greater than or equal to 1 mm. With an effective conversion efficiency of upwards of 40\%, we have laid a realistic avenue to substantially extend the brightness of photoinjector systems worldwide without major configuration changes and thus enhance current scientific capabilities on existing accelerators and reduce the cost of future accelerator facilities.

\begin{acknowledgments}
We thank Yuantao  Ding and Christopher Mayes at SLAC National Accelerator Laboratory for their helpful discussions.

This work is supported by the U.S. Department of Energy, Office of Science, Office of Basic Energy Sciences under Contract No. DE-AC02-76SF00515.
\end{acknowledgments}

\bibliography{bib}

\end{document}


\title{Supplemental Material: Temporal shaping of narrow-band picosecond pulses via non-colinear sum-frequency mixing of dispersion-controlled pulses}

\author{Randy Lemons}
 \email{rlemons@slac.stanford.edu}
\affiliation{%
 ~SLAC National Accelerator Laboratory and Stanford University, 2575 Sand Hill Rd, Menlo Park, CA 94025
}%
\affiliation{%
 ~Colorado School of Mines, 1500 Illinois St, Golden, CO 80401
}%
\author{Nicole Neveu}
 \email{nneveu@slac.stanford.edu}
\author{Joseph Duris}
\author{Agostino Marinelli}\affiliation{%
 ~SLAC National Accelerator Laboratory and Stanford University, 2575 Sand Hill Rd, Menlo Park, CA 94025
}%
\author{Charles Durfee}
\affiliation{%
 ~Colorado School of Mines, 1500 Illinois St, Golden, CO 80401
}%
\author{Sergio Carbajo}
 \email{scarbajo@slac.stanford.edu}
\affiliation{%
 ~SLAC National Accelerator Laboratory and Stanford University, 2575 Sand Hill Rd, Menlo Park, CA 94025
}%

\maketitle

\section{Control of pulse shape and duration}

\begin{figure}[htp!]
	\centering
	\includegraphics[width=0.7\linewidth]{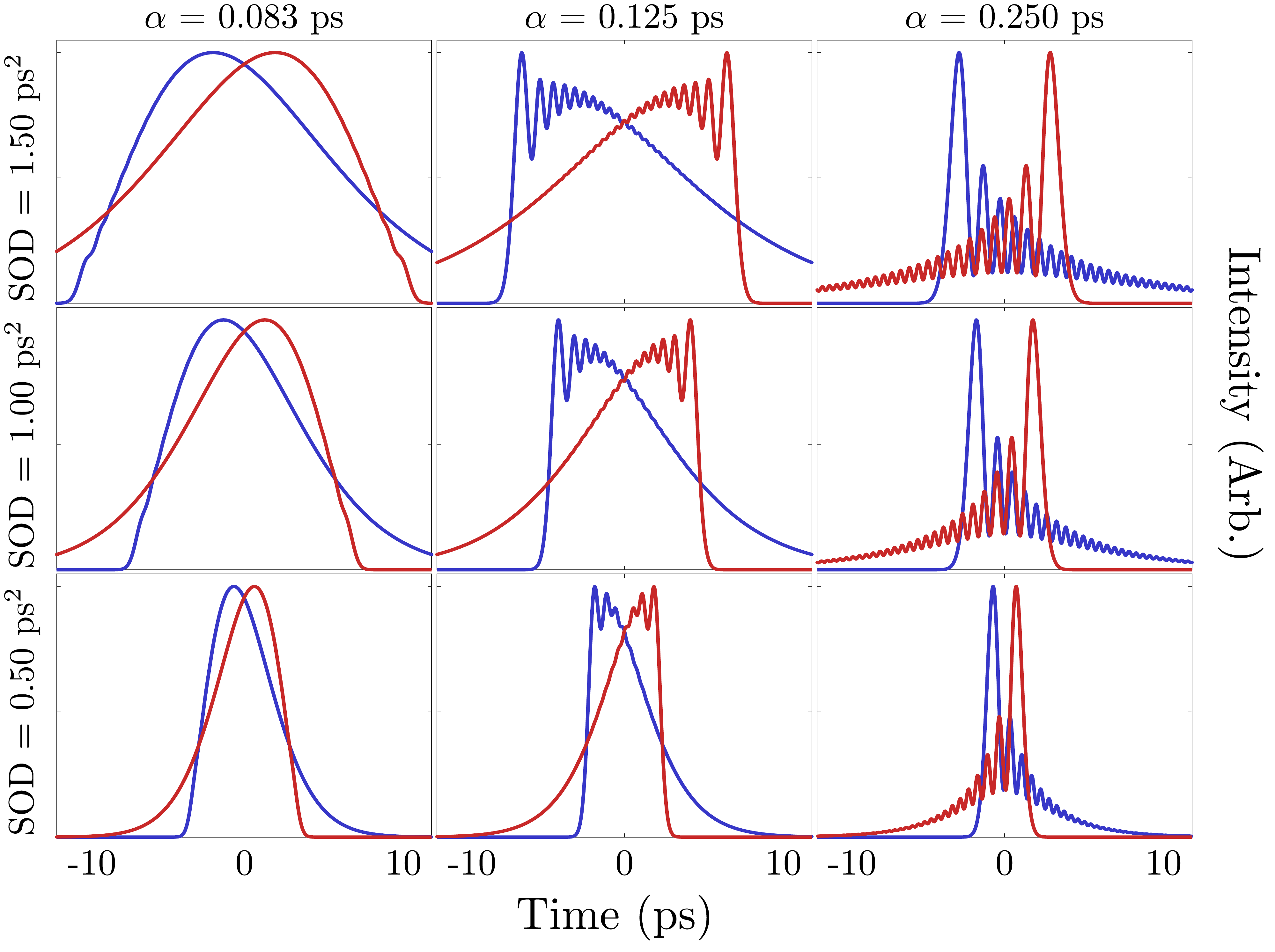}
	\caption{Pulse envelopes from different combinations of SOD and $\alpha$ demonstrating invariance of pulse shape to duration (row) instead being determined by the magnitude of $\alpha$ (column). The blue plot results from combinations with negative SOD and positive TOD while the red plot is positive SOD and negative TOD.}
  \label{FIG:3x3_Array_Alpha}
\end{figure}

The temporal intensity profile of a broadband laser pulse is heavily affected by the spectral phase of the constituent wavelengths. When representing the spectral phase as a Taylor expansion in frequency around the central frequency of the pulse
\begin{equation}
    \varphi(\omega) = \varphi_0 + \varphi_1 (\omega-\omega_0) + \frac{\varphi_2}{2!} (\omega-\omega_0)^2 + \frac{\varphi_3}{3!} (\omega-\omega_0)^3 + \frac{\varphi_4}{4!} (\omega-\omega_0)^4 + \dots,
\end{equation}
the effect on pulse shape is determined by the order of the term along with magnitude and sign. Here $\varphi_j$ is the $j^{th}$ derivative of $\varphi(\omega)$, evaluated at $\omega_0$. Second order dispersion (SOD), $\varphi_2$, is an even function around the central frequency and thus has a symmetric stretching effect to the temporal profile regardless of sign of the coefficient. However, the sign still determines whether the higher or lower frequency components lead the central frequency in time, with higher components leading for negative signs, and thus still has an effect when considering interactions with other dispersion orders. Third order dispersion (TOD), $\varphi_3$, is an odd function and results in a pre or post-pulse ringing depending on the sign of the coefficient with pre-pulses generated by positive sign. Higher orders have a similar even-odd-even pattern of effects but we do not include or cover their effects here.

In addition to the sign of the individual orders, the relative magnitudes also have a significant effect on the pulse shape. The trivial cases are pulses with some SOD and no TOD or visa versa. In these cases the pulse shape is fully representative of the appropriate orders' effect on temporal intensity. However, when varying amounts of SOD and TOD are added to the same pulse, the shape is determined instead by the relative amount of one to the other and thus we define a ratio,
\begin{equation*}
    \alpha = \frac{\varphi_3 / ps^3}{\varphi_2/ps^2},
\end{equation*}
to characterize pulse shape regardless of the absolute magnitude of either. $\alpha$ is defined such that value grows as TOD is added so that, at a glance, the magnitude of $\alpha$ represents departure from the typical Gaussian pulse shapes. Fig. \ref{FIG:3x3_Array_Alpha} demonstrates these properties of $\alpha$. Each column highlights the qualitative invariance to the magnitudes of the dispersion by locking the relative amounts. In this case TOD was calculated for a fixed $\alpha$ and changing SOD. Additionally , as $\alpha$ grows (left to right) the temporal intensity becomes less Gaussian due to a relatively larger amount of TOD for a given SOD.

\section{Direct calculation of UV profiles}

\begin{figure}[htp!]
	\centering
	\includegraphics[width=1\linewidth]{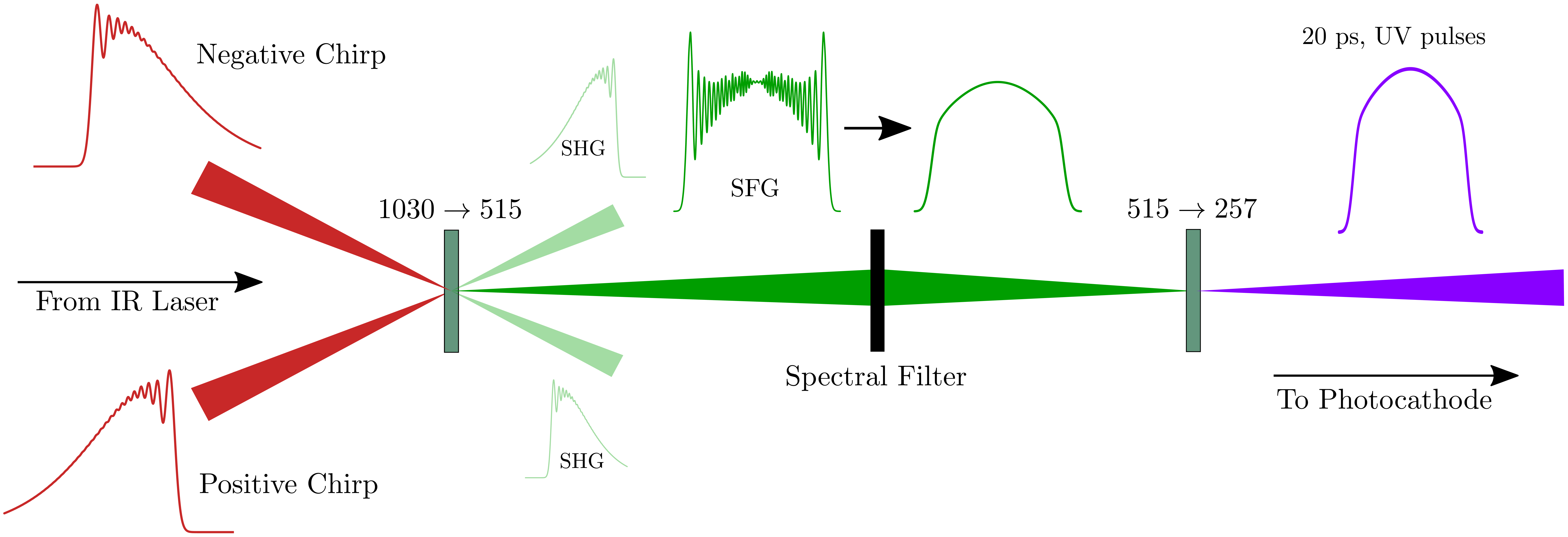}
	\caption{UV DCNS pulse generation from highly dispersed infrared pulses. The main stages are opposite chirp sum frequency generation (SFG) from 1030 nm to 515 nm, spectral filtering of SFG pulses, and finally second harmonic generation from 515 nm to 257 nm. The envelope representations shown here are in the time domain}
  \label{FIG:OpChirp_Full}
\end{figure}

A diagram of the ultraviolet (UV) dispersion controlled nonlinear synthesis (DCNS) pulse generation can be seen in Fig. \ref{FIG:OpChirp_Full}. The UV light is generated as a cascade of nonlinear conversion steps with spectral filtering taking place on the 515 nm light before the second conversion stage. Generating a UV profile from the 1030 nm light is involved and requires building the 1030 nm light from desired SOD and $\alpha$, simulating the nonlinear dynamics in the sum frequency generation (SFG) crystal using the methods in section \ref{SEC:optics}, applying an amplitude filter to the SFG light in the spectral domain and re-generating the temporal profile by inverse fourier transform, and finally simulating the nonlinear dynamics in the second crystal. For this process, generating the initial 1030 nm pulses and spectrally filtering the 515 nm light are computationally negligible compared to each of the split-step Fourier crystal simulations. As such, any computational speed increases to aid in the necessarily large amount of trials needed by a genetic algorithm would be found in the crystal simulations. To this end, eliminating one of the crystal simulations would essentially half the computational time of the entire optical pulse simulation.

Second harmonic generation is governed by the coupled equations,
\begin{subequations}
    \begin{align}
        \frac{d A_1}{d z} &= \frac{2 i d_{eff} \omega_1^2}{k_1 c^2}A_1^* A_2 e^{i \Delta k z} \\
        \frac{d A_2}{d z} &= \frac{i d_{eff} \omega_2^2}{k_2 c^2}A_1^2 e^{i \Delta k z}
    \end{align}
    \label{EQN:SHGEqns}
\end{subequations}
where $A_1$ is the incident field to the crystal, $A_2$, $\Delta k = 2k_1-k_2$, $d_{eff}$ is the nonlinear coefficient of the crystal, and $\omega_i$ and $k_i$ are the central frequency and wave vector of their respective fields~\cite{boyd2019nonlinear}. In the DCNS process, the second harmonic conversion from 515 nm to 257 nm is a remarkably simple situation mathematically. The 515 nm light is spectrally narrow with very little phase variation and as such $\Delta k$ is very small. In this case, the field amplitude of the second harmonic field is well approximated as a scaled version of the incident lights' intensity profile.

\begin{figure}[htp!]
	\centering
	\includegraphics[width=0.7\linewidth]{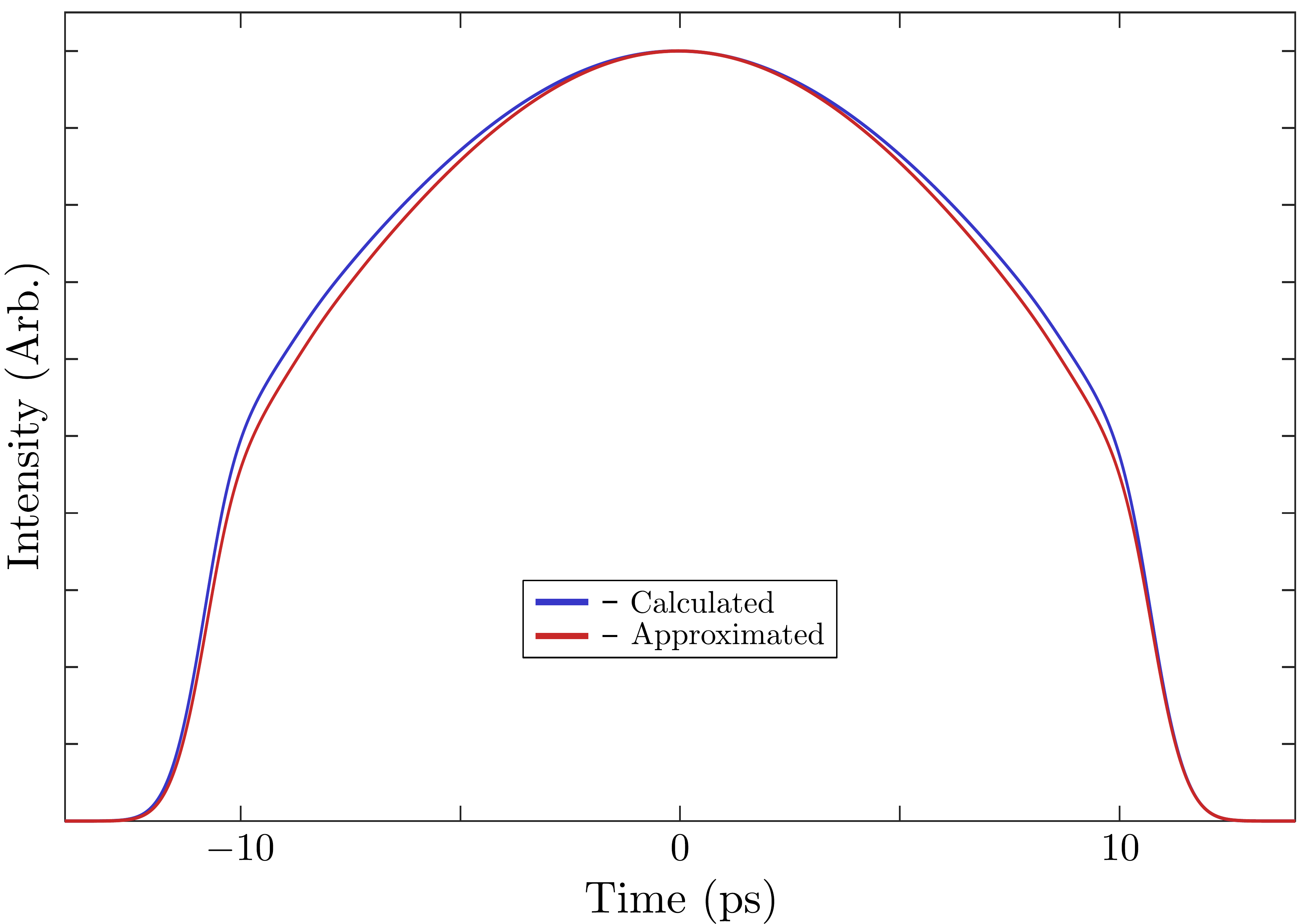}
	\caption{Generation of UV intensity profile for the 0.5 nm filter optimal parameters (SOD: $\pm 3.436$, $\alpha$: -0.128) via full crystal simulation and approximation by squaring the 0.5 nm filtered 515 nm light. The full simulation intensity profile (blue) differs from direct calculation (red) by 0.83\% root mean squared error and while requiring significantly more computation time.}
  \label{FIG:Square_515}
\end{figure}

Fig. \ref{FIG:Square_515} displays the UV temporal intensity profile calculated first by running a full simulation of the crystal dynamics (blue) and then approximated by assuming the UV field is given by the intensity profile of the 515 nm light (red). The approximated field in this case displays all of the characteristics of the calculated field including the residual ringing on the sides of the pulse in addition to having a nearly identical full width at half maximum (FWHM). The root mean squared error between the approximated intensity profile and the calculated one is only 0.83\%.  This approximation would cease to be valid in the case where the second harmonic field starts converting energy back to the incident field; however, this is easily mitigated in a real world situation by lowering the intensity in the crystal or using decreasing crystal length.

By using this approximation rather than the full simulation we were able to reduce the computational complexity of the problem and save nearly a full day of processing time for less than 1\% of difference.

\section{Pulse shape tolerance to variance of $\alpha$ and filter width}

\begin{figure}[htp!]
	\centering
	\includegraphics[width=0.7\linewidth]{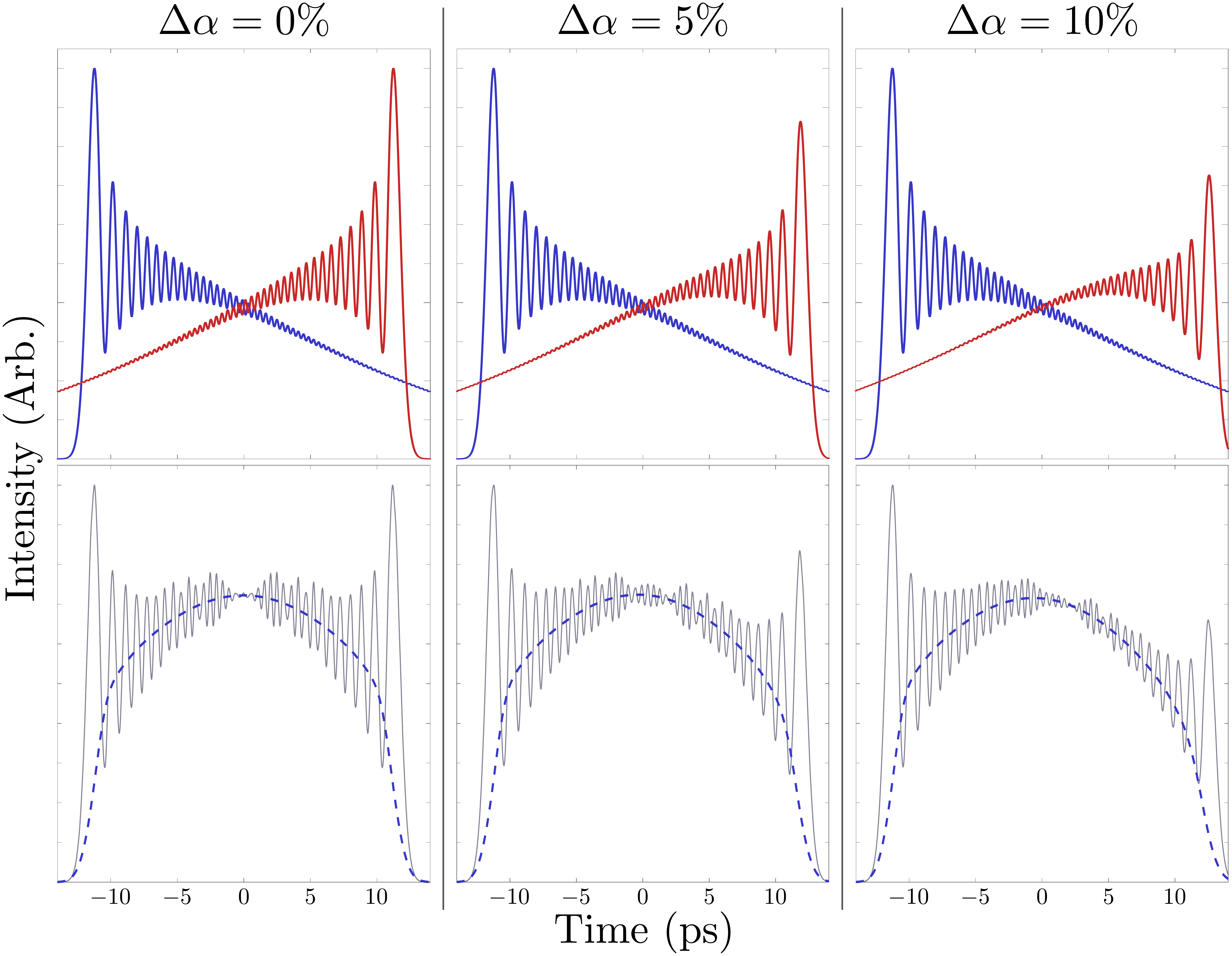}
	\caption{Pulse shape changes of the 0.5 nm filter 515 nm pulse with the GA optimized parameters (SOD: $\pm 3.436$, $\alpha$: -0.128) to percent variations in $\alpha$ on one input pulse to mimic real world variance and experimental imperfections. As the variance in $\alpha$ is increased (left to right) the side oscillations are asymmetrically changed along with the slope of the filtered pulse.}
  \label{FIG:3x6_Alpha_Var}
\end{figure}

\begin{figure}[htp!]
	\centering
	\includegraphics[width=0.7\linewidth]{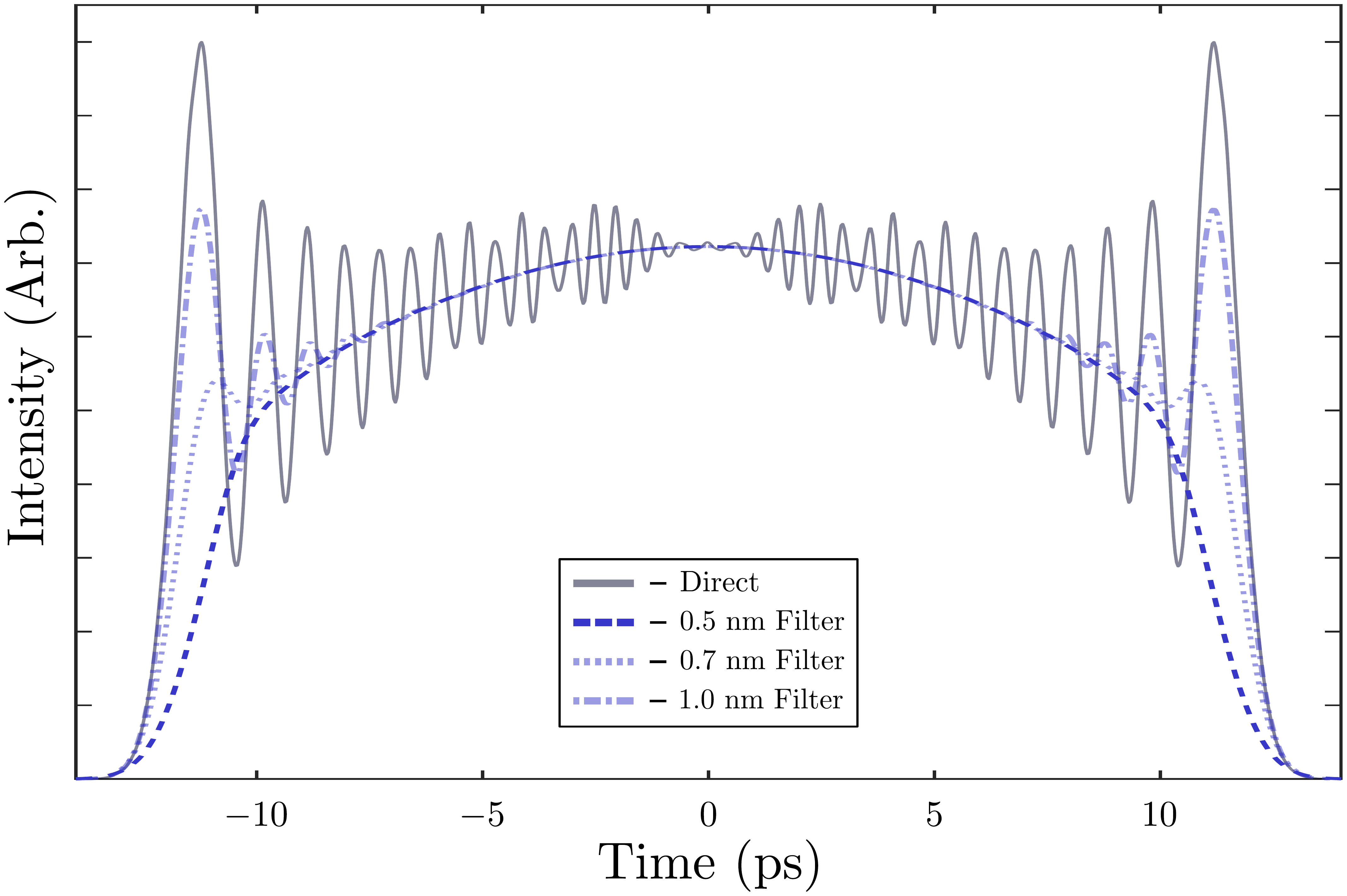}
	\caption{Pulse shape changes of the 0.5 nm filter 515 nm pulse with the GA optimized parameters (SOD: $\pm 3.436$, $\alpha$: -0.128) to changing filter widths. As the filter widths increase the side oscillations are not as attenuated due to increase high frequency contributions.}
  \label{FIG:Filt_Changes}
\end{figure}

As demonstrated in Sec. 1, $\alpha$ is the dominent parameter on controlling the presence of side oscillations in the mixing process with larger $\alpha$ indicating more TOD and more oscillations. When both input pulses experience a uniform variance in $\alpha$ the resulting mixed pulse shape has the side oscillations uniformly changed, based on the sign of the change in $\alpha$. However, when a single input pulse experiences a change in $\alpha$ that is unmatched on the other input pulse, the mixed pulse begins to show a loss of symmetry. Variations of this nature would be nearly unavoidable in a real world implementation of our proposed method. 

As seen in Fig.\ref{FIG:3x6_Alpha_Var}, the unfiltered profile of a mixed pulse has a drastically changed profile for even a 5\% decrease in $\alpha$ of one input pulse with respect to the other and this change is further exasurbated when the difference is increased to 10\%. When these profiles are then filtered the difference in pulse shape is reduced to the point that the 5\% difference displays a very similar profile. The filtered 10\% difference case shows a reduction in the effect but still results in an asymmetric shape that is likely to affect an intended application. It is worth noting that this asymmetric degradation is seemly linear and might result in a tunable parameter between square pulse and triangular pulse shapes, should TOD be independently controllable.

Another parameter that has and effect on the resulting pulse shape is the width of the filter used after mixed pulse synthesis. Like $\alpha$, a wider filter will generally result in a greater retention of the side band oscillations. 

In Fig \ref{FIG:Filt_Changes} we have generated a mixed pulse using the GA optimal parameters for a 0.5 nm filter (SOD: $\pm 3.436$, $\alpha$: -0.128) and then change the filter width while maintaining SOD and . This results in the intuitive change that larger filters retain more of the high frequency components that interfere with the main spectral lobe and thus retain more side oscillations. While the SOD and $\alpha$ values used are not identical to the GA optimal parameters for 0.7 nm and 1.0 nm filters, they are similar enough to remain a valid qualitative comparison to the changing of filter widths.

\bibliography{bib}